\documentclass[12pt]{iopart}
\newcommand{\om}{\Omega_{\rm M}}
\newcommand{\ola}{\Omega_{\rm\Lambda}}
\usepackage{iopams}
\usepackage{epsfig} 
\usepackage{citesort}
\begin{document}

\title{Strong lensing, cosmology and lensing halos}
\author{Edvard M\"{o}rtsell and Christoffer Sunesson}
\address{Department of Astronomy, Stockholm University, SE-106 91 Stockholm, Sweden}
\eads{\mailto{edvard@astro.su.se, sunesson@astro.su.se}}

\date{{\today}}

\begin{abstract}
With future wide and deep cosmological sky surveys, a large number of
gravitationally lensed, multiply imaged systems will be found. In
addition to multiply imaged galaxies and quasars, sources will include
transient events like supernovae and gamma ray bursts in which case
very accurate time delay measurements are possible. Also, large
numbers of systems with several lensed sources behind a single lens
will be observed. In this paper, we review and compare different
possibilities of using future strong lensing data to probe lens matter
distributions and to determine the Hubble parameter and the matter
density of the universe. Specifically, we investigate the possibility
to break the well-known degeneracy between dark matter halo profiles
and the Hubble parameter using observed flux ratios. We also
investigate how strong lensing can provide useful constraints on the
matter density of the universe independently of the flux ratios and
other cosmological probes.
\end{abstract}


\section{Introduction}\label{sec:intro}
In the last decade, a picture of our universe being dominated by dark
energy and dark matter has emerged. Elements of this picture include
observations of Type Ia supernova (SNIa) distances
\cite{Riess1998,Perlmutter1999,Knop2003,Riess2004}, anisotropies in the cosmic
microwave background (CMB)
\cite{Spergel2003,Hanany2000,Netterfield2002,Sievers2003,Kovac2002,Kuo2004}
and the large scale structure (LSS) of galaxies
\cite{Tegmark2004,Cole2005}. 
For a recent concise review of cosmological bounds on dark matter and
dark energy, see Ref.~\cite{Hannestad2005}. A flat universe with
$\ola\sim 0.7$ and $\om\sim 0.3$ provides a good fit to all current
observations, indicating a recent ($z\sim 0.5$) transition from a
decelerating universal expansion to an accelerating universal
expansion. The current expansion rate (and hence overall scale) of the
universe as measured by the Hubble Space Telescope (HST) Key Project
\cite{Freedman2001} is $h=0.72\pm 0.08$ where $h$ is the dimensionless
Hubble constant, $h=H_0/(100 \,{\rm km \,s^{-1} \,Mpc^{-1}})$.

Though the picture described above is very effective in explaining all
observational data with just a few free parameters, the nature of the
dark energy and dark matter is still unknown. The simplest dark energy
model is the cosmological constant. Difficulties in theoretically
explaining the magnitude of the energy density has brought about a
large number of alternative explanations, including quintessence
\cite{Ratra1988,Wetterich1988,Peebles1988}, k-essence 
\cite{Chiba2000,Armendariz-Picon1999} and phantom energy 
\cite{Caldwell2002,Carroll2003}. However, current 
observations are still best explained by the cosmological constant
\cite{Hannestad2002,Hannestad2004}.

On large scales, dark matter is well-described as being
non-relativistic and collisionless, i.e.~cold dark matter (CDM). On
galaxy scales however, there has been a dispute whether there is a
discrepancy between numerical simulations of galaxy formation in CDM
models
\cite{Navarro1997,Ghigna2000,Navarro2004,Fukushige2004,Graham2005} and
observations of rotation curves of dark matter dominated galaxies
\cite{vandenBosch2001,deBlok2001,Simon2005}. Precise observational 
determinations of halo density profiles will help in understanding
whether there is any contradiction with the simulations and provide
further insight into the properties of dark matter.

When investigating the matter distribution of the universe,
gravitational lensing has the advantage of being equally sensitive to
all kinds of matter, regardless of it's microscopic properties
(e.g.~being baryonic or non-baryonic) and dynamical state. Both strong
lensing (multiply imaged sources) and weak lensing (weakly distorted
source images) has been used to constrain the matter distribution in
galaxy scale \cite{Cohn2001,Davis2003,Brainerd2003} and cluster scale
\cite{Gavazzi2003,Broadhurst2005a,Natarajan2002,Dahle2003} halos.

Besides being able to provide information on the matter distribution
of dark matter halos, gravitational lensing has also the potential to
constrain cosmological parameters such as $\om,\ola$ and $h$.  More
than 40 years ago, Refsdal showed how the time delay between multiple
images of supernovae (SNe) could be used to measure $h$ and the mass
of the lensing galaxy \cite{Refsdal1964}. To date -- for galaxy mass
lenses -- approximately 100 multiply imaged sources (including
galaxies and quasars) have been observed out of which there are eleven
well-determined time delay measurements (all quasars)
\cite{Kochanek2004,Davis2003,York2005}.
For several reasons, it has been difficult to implement Refsdal's
method. First, time delays between quasar images are notoriously
difficult to measure compared to what would be the case for transient
sources like supernovae. Second, there is a degeneracy between the
derived value of $h$ and the matter distribution of the lensing
galaxy. Modeling the lenses as more concentrated gives larger values
of $h$ and vice versa \cite{Kochanek2002a,Kochanek2003}.

Weak lensing can be used to probe the matter distribution of the
universe also on very large scales. For reviews on the current state
of cosmological constraints from LSS weak lensing, see
Refs.~\cite{Refregier2003,Hoekstra2002,Mellier2002}. There has also
been attempts to constrain $\ola$ through the statistics of
gravitationally lensed quasars with disparate results
\cite{Kochanek1996,Falco1998,Chiba1999,Keeton2002}.

Future strong lensing data will include observations from
e.g.~Pan-STARRS\footnote{\tt http://www.pan-starrs.org}, the Large
Synoptic Survey Telescope (LSST)\footnote{\tt http://www.lsst.org},
the Joint Dark Energy Mission (JDEM) contender the Supernova
Acceleration Probe (SNAP)\footnote{\tt http://snap.lbl.gov} and the
James Webb Space Telescope (JWST)\footnote{\tt
http://www.jwst.nasa.gov}. In this note, we use simulated strong
lensing data for a SNAP like mission combined with follow up
observations from, e.g. JWST to investigate and compare different
possibilities of constraining the slope of lens density profiles, the
Hubble parameter and the matter density of the universe.

In Sec.~\ref{sec:model} we present our lens model. In
Sec.~\ref{sec:imis}, we investigate different uses of a data set
consisting of a large number of lens systems, each with a multiply
imaged source with well-determined time delay. This section extends
and generalizes earlier work in Refs.~\cite{Goobar2002} and
\cite{Mortsell2005} (see also 
Refs.~\cite{Holz2001,Oguri2003a,Oguri2003b,Bolton2003}). In
Sec.~\ref{sec:mss}, we compare our results with what can be obtained
by studying lens systems with several multiply lensed sources in a
single lens (cf. the case of Abell 1689
\cite{Broadhurst2005a}), with or without measured time delays.
Finally, Sec.~\ref{sec:summary} contains a summary and a discussion of
our results.

The cosmology used in the simulations throughout this paper is
$h=0.65$, $\om =0.3$ and $\ola =0.7$. Our default dark matter halo is
the singular isothermal sphere described in Sec.~\ref{sec:model}.

\section{Lens model}\label{sec:model}
The lens equation relating the angular source position $\vec\theta$
and the angular image position $\vec\beta$ is given by
\cite{Schneider1992}
\begin{equation}\label{eq:lenseq1}
   	\vec\beta D_s=\vec\theta D_s-\hat{\vec\alpha}D_{ls},
\end{equation}
where $\hat{\vec \alpha}$ is the light deflection angle and $D_{ls},
D_l$ and $D_s$ are angular diameter distances between source and lens,
lens and observer, and source and observer, respectively.  The lens
equation can be rewritten in dimensionless form by introducing an
arbitrary length scale $\xi_0$ in the lens plane and a corresponding
length scale $\eta_0=\xi_0D_s/D_d$ in the source plane. In terms of
the dimensionless vectors $\vec x=\vec\theta D_l/\xi_0$ and $\vec
y=\vec\beta D_s /\eta_0$ the lens equation takes the form
\begin{equation}\label{eq:lenseq3}
  	\vec y=\vec x -\vec\alpha(\vec x),
\end{equation}
where 
\begin{equation}\label{eq:sda} 
  	\vec\alpha(\vec x)=\frac{D_d D_{ls}}{D_s \xi_0}\hat{\vec
  	\alpha}(\vec\xi_0 \vec x)
\end{equation}
is the scaled defelction angle.

In this paper, we assume that the matter distribution in lenses can be
described by a simple power-law density profile
\begin{equation}
	\rho(r) \propto r^{-\eta}.
\end{equation}
Since almost all images of multiply lensed sources are at a limited
range of small $r$, this is a good approximation despite the fact that
the slope is expected to change with radius at large $r$
\cite{Navarro1997,Ghigna2000,Navarro2004,Fukushige2004}. Note that the
baryonic contribution to the density profile of lensing galaxies is
non-negligible at small radii and that results from strong lensing not
necessarily reflect the nature of pure dark matter halos.

For $\eta =2$, we obtain the familiar singular isothermal sphere (SIS)
model built upon the assumption that the mass components behave like
particles in an ideal gas, confined by their spherically symmetric
gravitational potential;
\begin{equation}
  \label{eq:sis}
  \rho_{SIS} (r)=\frac{v^2}{2\pi G}\frac{1}{r^2}.
\end{equation}
Here, $v$ is the line-of-sight velocity dispersion of the mass particles.

For all lensing purposes in this paper, matter distributions can be
described by the projected Newtonian gravitational potential
\begin{equation}
  	\Psi =\frac{2D_lD_{ls}}{c^2D_s\xi_0^2}\int\Phi dl.
\end{equation}
For the power-law density profile, the projected potential is given by
\cite{Kochanek2004}
\begin{equation}\label{eq:psi}
  	  \Psi\propto x^{3-\eta}
\end{equation}
where $x$ is the impact parameter (in arbitrary units, $\xi_0$),
i.e.~the minimum distance between the light ray and the lens centre.

\subsection{The Einstein radius}
Since the scaled deflection angle $\vec\alpha =\nabla\Psi$, for a
spherically symmetric lens we have
\begin{equation}\label{eq:psi2}
    	\Psi =\frac{x_E^{\eta -1}}{3-\eta}x^{3-\eta},
\end{equation}
where $x_E$ is the Einstein radius; the solution to the lens equation
for $y=0$. The lens equation for the power-law lens can then be
written (where $x>0$ and $\alpha$ is directed towards the center of
the lens)
\begin{equation}\label{eq:le}
    	y=x-\alpha =x-x_E^{\eta -1}x^{2-\eta}.
\end{equation}
For multiple images, we have
\begin{equation}\label{eq:le2}
    	y=x_1-\alpha_1 =\alpha_2-x_2
\end{equation}
and 
\begin{equation}\label{eq:x_E}
    	x_E=\left[\frac{x_1+x_2}{x_1^{2-\eta}+
    	x_2^{2-\eta}}\right]^{\frac{1}{\eta -1}}.
\end{equation}
Putting $\xi_0=D_l$, i.e.~denoting positions in terms of angles, we
get 
\begin{equation}\label{eq:theta_E}
    	\theta_E=2\left[\frac{(\theta_1+\theta_2)^{2-\eta}}{\theta_1^{2-\eta}+
    	\theta_2^{2-\eta}}\right]^{\frac{1}{\eta-1}} \theta_{\rm SIS},
\end{equation}
where $\theta_{\rm SIS}$ is the Einstein radius for a SIS halo,
\begin{equation}\label{eq:theta_SIS}
    	\theta_{\rm SIS}=4\pi\left(\frac{v}{c}\right)^2\frac{D_{ls}}{D_s}.
\end{equation}

\subsection{Time delay}
The time delay for a gravitationally lensed image as compared to an
undeflected image is in the general case given by \cite{Schneider1992}
\begin{equation}\label{eq:T}
    	T=\frac{\xi_0^2D_s}{D_lD_{ls}}(1+z_l) \left[\frac{({\bf x}
    	-{\bf y})^2}{2}-\Psi \right].
\end{equation}
The time delay between two lensed images is given by $\Delta
t=T_2-T_1$. For an isothermal lens, we have 
\begin{equation}
    	\Delta t_{\rm SIS}=\frac{1}{2}\frac{D_lD_s}{D_{ls}}(1+z_l)
    	(\theta_1^2-\theta_2^2)=2\hat\alpha D_l(1+z_l)\beta .
\end{equation}
Note that the time delay is independent of the source redshift. In the
general case of $\eta\ne 2$, we can rewrite Eq.~(\ref{eq:T}) in terms
of
$q\equiv\Delta\theta/\hspace{-.1cm}<\hspace{-.1cm}\theta\hspace{-.1cm}>$
where $\Delta\theta =\theta_1-\theta_2$ and
$<\hspace{-.1cm}\theta\hspace{-.1cm}>=(\theta_1+\theta_2)/2$ and
Taylor expand to obtain (see also Refs.~\cite{Kochanek2004} and
\cite{Chang1977})
\begin{eqnarray}\label{eq:delt}
    	\Delta t&=&\frac{(\eta -1)}{2}\frac{D_lD_s}{D_{ls}}(1+z_l)
    	(\theta_1^2-\theta_2^2)\left[1-\frac{(2-\eta)^2}{12}
    	q^2+\mathcal{O} (q^3)\right]\nonumber\\ &\simeq&(\eta
    	-1)\Delta t_{\rm SIS}\left[1-\frac{(2-\eta)^2}{12} q^2\right].
\end{eqnarray}
There is a strong degeneracy between $\eta -1$ and $h$ (which comes in
through the distances). This degeneracy can be broken by including
data from the observed image luminosities; the flux ratio.

\subsection{Flux ratio}
We denote the flux ratio $r\equiv |\mu_1/\mu_2|$. The magnification
$\mu$ for a source at position $y$ observed at position $x$ is given
by
\begin{equation}
    	|\mu|=\left|\frac{x}{y}\frac{dx}{dy}\right|.
\end{equation}
For a power-law lens, we can express the flux ratio as
\begin{eqnarray}\label{eq:r}
    	r&=&\left|\frac{\theta_1}{\theta_2}\right|\left|
    	\frac{1-(2-\eta)\theta_2^{1-\eta}(\theta_1+\theta_2)/(\theta_1^{2-\eta}+\theta_2^{2-\eta})}
    	{1-(2-\eta)\theta_1^{1-\eta}(\theta_1+\theta_2)/(\theta_1^{2-\eta}+\theta_2^{2-\eta})}\right|\nonumber\\
    	&=&r_{\rm SIS}\left|
    	\frac{1-(2-\eta)\theta_2^{1-\eta}(\theta_1+\theta_2)/(\theta_1^{2-\eta}+\theta_2^{2-\eta})}
    	{1-(2-\eta)\theta_1^{1-\eta}(\theta_1+\theta_2)/(\theta_1^{2-\eta}+\theta_2^{2-\eta})}\right|.
\end{eqnarray}
Expanding this to second order in $q$, we obtain
\begin{equation}\label{eq:rq}
	r\simeq 1+(\eta-1)q + \frac{1}{2}(\eta-1)^2 q^2 + {\cal
	O}(q^3).
\end{equation}
Note that the flux ratio is independent of the value of $h$.

\section{Individual multiple image systems}\label{sec:imis}
With a SNAP like JDEM contender, a large number of core collapse
supernovae (CC SNe) will be discovered \cite{Goobar2002a}. Monte Carlo
simulations performed using the SNOC package \cite{Goobar2002b}
predicts that out of a total of $\sim 10^6$ CC SNe at $z<5$ in a 20
square degree field during three years \cite{Dahlen1999}, $\sim 850$
will be multiply imaged with an I-band or J-band peak brightness
$<28.5$ mag for the dimmest image\footnote{Since these simulations
were performed, the SNAP specifications has been slightly degraded and
the corresponding point-source magnitude limit is now 27.7. This
number would yield $\sim 400$ multiply imaged CC SNe.}. We demand that
the surface brightness of the lens galaxy is fainter than 24 I
magnitudes per square arcsecond at the position of the faintest image,
in order to avoid contamination from the lens galaxy. In accordance
with Ref.~\cite{Mortsell2005}, we also introduce a quality factor,
$f$, giving the fraction of the lens systems that are ``simple'',
i.e.~a single dominant deflector with moderate ellipticity and
external shear. In the following, we assume a quality factor, $f=0.5$,
in accordance with currently observed systems
\cite{Kochanek2004}. This cut leaves a total of $\sim 400$ systems. 

Here, we investigate how measurements of the image positions, time
delays, flux ratios and lens galaxy properties for such systems could be
used to constrain the slope of lens density profiles, $\eta$, the
Hubble parameter, $h$, and the matter density of the universe, $\om$.

\subsection{Error budget}\label{sec:errors}
Our error assumptions basically follow Refs.~\cite{Goobar2002a} and
\cite{Mortsell2005}. We conservatively use a positional uncertainty of
$\sigma_{\theta} =0.01''$. Combining data points from several filters
could be combined to give a time delay uncertainty $\sigma_{\Delta t}$
of 0.05 days for the SNAP mission. However, this number may be
degraded if the lightcurve is modified by microlensing. In the
following we set $\sigma_{\Delta t} = 0.15$ days. For the flux ratios,
we assume the error $\sigma_r$ to be dominated by microlensing
\cite{Schneider1987,Wozniak2000} and lensing by CDM substructure
\cite{Dalal2002,Keeton2003,Mao2004}. An error estimate of $\sigma_r/r
= 0.5$ is employed. Ideally, one would like to use the proper
probability distribution functions for the microlensing magnification
probability as a function of, e.g. position in lens galaxy. In the
following however, we assume a simple gaussian distribution and assume
that extreme events with very large microlensing magnification factors
can be identified from, e.g. pecularities of the lightcurves and be
removed from the sample. This will however only be true if the
microlensing timescale is smaller than the timescale of the intrinsic
source variations. Redshifts error are negligible; we adopt
$\sigma_{z_l} = 0.001$ from photometric redshift measurements based on
SNAP multi-band photometry.

Even though we can derive lensing statistics using simple spherical
galaxy profiles, it is important to include lens ellipticities and
galaxy environments in the analysis \cite{Keeton2004}. Motivated by
models of known gravitational lens systems, we
assume an uncertainty in the alignment between the mass distribution
and the optical light distribution of $\langle \Delta\phi^2
\rangle^{1/2} < 10$ degrees \cite{Kochanek2002b}. 
From N-body simulations and semianalytic models of galaxy formation,
we can estimate the levels of external shear due to structure near the
lens in gravitational lens systems to be $\gamma_{\rm ext} = 0.058$,
with an rms dispersion of 0.071 \cite{Holder2003}. We modify our
simple simulated lensing systems to include this effect. We also add
an uncertainty due to scatter in the slope of individual halos of
$\Delta\eta =0.2$.

The addition of a constant surface density $\kappa_c$ to a lens,
changes the time delay by a factor $1-\kappa_c$ but leaves the image
positions and flux ratios unchanged. In the following we have assumed
the effects from external convergence to be comparable to or smaller
than the effects from external shear and a scatter in $\eta$. However,
it should be noted that the effect potentially can be very important
\cite{Bar-Kana1996,Oguri2005}. The error in the derived value of $h$ 
will be directly proportional to the systematic error in the assumed
value of the external shear.

When constraining $h$, we use a prior on the matter density of
$\Omega_m = 0.30 \pm 0.04$ \cite{Tegmark2004}. Results are very
insensitve to the exact value and size of the error on this prior.

\subsection{The Hubble parameter}\label{sec:hubble}
\begin{figure}
  	\begin{center} \epsfxsize=\textwidth \epsffile{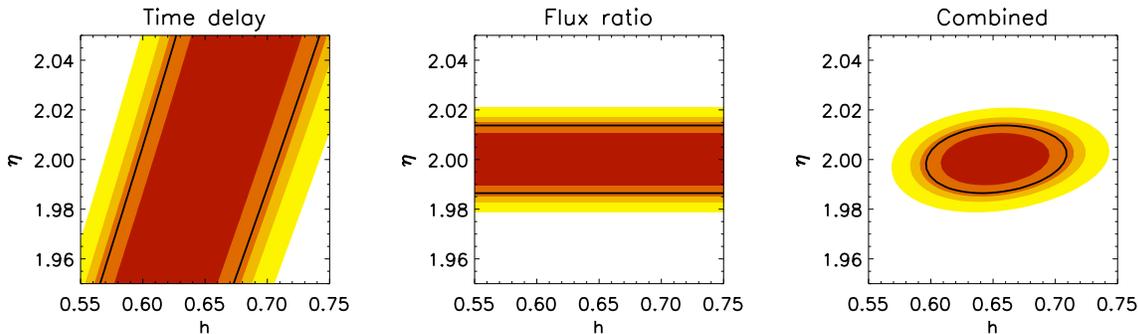}
  	\caption{Constraints in the $[\eta,h]$-plane using $\sim 400$
  	multiply imaged SNe (corresponding to a quality factor
  	$f=0.5$). In the left panel, constraints from the time delays
  	are shown, in the middle panel constraints from the flux
  	ratios. Combined constraints are shown in the right
  	panel. Contours correspond to 68.3\,\%, 90\,\%, 95\,\% and
  	99\,\% confidence levels for two parameters. The black line
  	indicates the 95\,\% confidence level ($2\sigma$) for one
  	parameter.} \label{fig:at1} \end{center}
\end{figure}

For constraints in the $[\eta,h]$-plane, we follow the same recipe as
Ref.~\cite{Mortsell2005} and obtain very similar results, see
Fig.~\ref{fig:at1}. Small differences are due to the random
fluctuations in the simulation of lens systems as well as the fact
that we only include the scatter in the slope $\eta$ in the
uncertainty -- not as a modification of the simulated sample as was
the case in Ref.~\cite{Mortsell2005}. This is to ensure that our
confidence contours are centred at the correct value to facilitate
estimates of the contour sizes. Constraints from the time delays are
shown in the left panel and constraints from the flux ratios in the
middle panel. In deriving these constraints, we have used
Eqns.~(\ref{eq:delt}) and (\ref{eq:rq}) generalized to include the
effects from galaxy ellipticities and external shear, see Appendix
A. Contours correspond to 68.3\,\%, 90\,\%, 95\,\% and 99\,\%
confidence levels for both parameters to have values within their
borders. The black line indicates the 95\,\% confidence level for one
of the parameters to lie within the contour (i.e.~$2\sigma$). Because
of the strong degeneracy in the time delay between $h$ and $\eta$,
flux ratio measurements are needed to constrain the Hubble parameter.
Combining the results from the time delay and flux ratio measurements,
we are able to make a determination of $h$ within 10\,\% and --
perhaps more interestingly -- to determine $\eta$ at the per cent
level at 95\,\% confidence (right panel). Even for an extremely
conservative value of $f=0.05$, we can determine $\eta$ within 2\,\%
and $h$ within 25\,\%. It should be noted however that these numbers
may be severly degraded in the presence of systems with very large
microlens magnifications.

\subsection{The matter density}\label{sec:matter}
\begin{figure}
  	\begin{center} \epsfxsize=\textwidth \epsffile{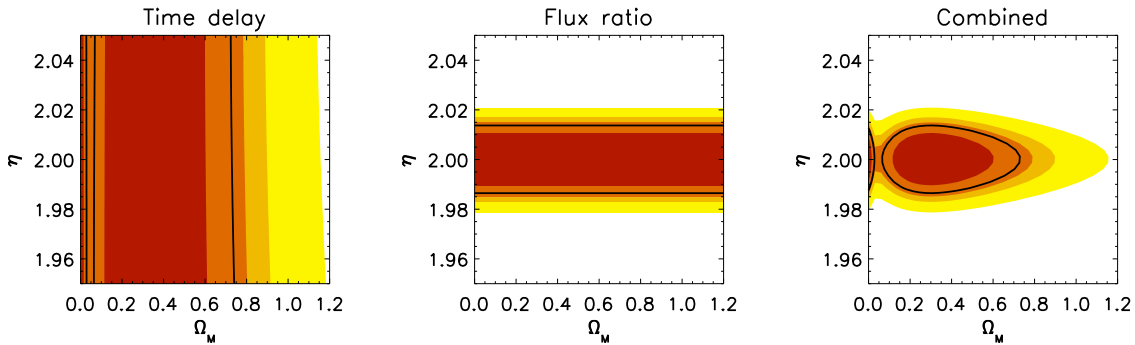}
  	\caption{Constraints in the $[\eta,\om]$-plane using $\sim
  	400$ multiply imaged SNe (corresponding to a quality factor
  	$f=0.5$). In the left panel, constraints from the time delay
  	ratios are shown, in the middle panel constraints from the
  	flux ratios. Combined constraints are shown in the right
  	panel.}\label{fig:a2} \end{center}
\end{figure}

Since the time delay is proportional to $h$ through the distances, we
can study the ratio of time delays in order to factor out the
Hubble constant and constrain $\om$ (assuming a flat universe). I
doing this, we wish to study ratios of lens systems with large
differences in redshift in order to maximize the sensitivity of the
ratio to the value of the matter density. Denoting $k_l=\Delta t_m /
\Delta t_n$, where $m$ and $n$ numbers the lens systems 
in the total sample, we want to combine the time delays (i.e. choose
$m$ and $n$) in order to maximize $(dk/d\om )/\sigma_k\equiv
k'/\sigma_k$ (i.e. maximizing the sensitivity to $\om$). This is
equivalent to maximizing the quantity
\begin{equation}
	\frac{\Delta t'_m / \Delta t_m - \Delta t'_n / \Delta
	t_n}{\sqrt{\Delta t_m^{-2}+\Delta t_n^{-2}}}.
\end{equation}
From a sample of $N$ lens systems, we are thus able to obtain $N-1$
time delay ratios with maximum sensitivity to $\om$. Note the
similarity of the method to the use of cross-correlation tomography in
weak lensing \cite{Jain2003}. The time delay is thus used to constrain
ratios of the distance combination $(D_lD_s)/D_{ls}$, see
Eq.~(\ref{eq:delt}). The flux ratio data is used to constrain $\eta$
as described in Sec.~\ref{sec:hubble}. In Fig.~\ref{fig:a2}, results
from the time delays (left panel), flux ratios (middle panel) and the
combined results (right panel) are shown. Note that the time delay
ratios basically constrain $\om$ independent of the value of $\eta$
which allows us to constrain the matter density independently of the
flux ratio measurements that may be plagued by large
uncertainties. The ``false'' minima at very low $\om$ is due to the
fact that the ratio of distances often is degenerate in $\om$. The
same effect can be seen in Fig.~\ref{fig:einstein} that basically
probes the distance combination $D_{ls}/D_s$, see
Eqns.~(\ref{eq:theta_E}) and (\ref{eq:theta_SIS}). Though not
comparable in precision to other probes of the matter density such as
the LSS of galaxies, strong lensing should be able to rule out an
Einstein-de Sitter universe with $\om =1$ at very high confidence
level and provide an independent sanity check of the matter
density. Setting $f=0.05$ leaves $\om$ more or less unconstrained.

\section{Multiple source systems}\label{sec:mss}
We next compare our results obtained with single source systems with
the possibilty to constrain the matter density in the universe and the
matter distribution in lenses using systems with multiple sources
being lensed by the same lens. The cross-section for multiple imaging
is larger for cluster mass lenses than for galaxy size halos. However,
the time delays for cluster lenses are generally very large. Also, if
the sources are galaxies, we are not able to measure the time delays.
Examples of such systems are Abell 2218 \cite{Soucail2004} and Abell
1689 where 30 background galaxies are being lensed by the cluster lens
and displays more than 100 images
\cite{Broadhurst2005a}.

The cross-section for multiple imaging by a SIS halo is given by
\begin{equation}
  \label{eq:micross}
  \hat\sigma_{mi}=16\pi^3\left(\frac{v}{c}\right)^4D_{ls}^2\ ,
\end{equation}
and the time delay for the two images is
\begin{equation}
  \label{eq:sistd}
  c\Delta t=32\pi^2\left (\frac{v}{c}\right )^4
            \frac{D_dD_{ls}}{D_s}(1+z_l)\frac{r-1}{r+1}
\end{equation}
where $r$ is the flux ratio. For a lens redshift of $z_l=0.5$ and a
source redshift $z_s=1.5$, we get
\begin{equation}\label{eq:deltr}
  \Delta t\sim 2\left (\frac{v}{200\,{\rm km/s}}\right )^4
	\frac{r-1}{r+1}\;{\rm months}.
\end{equation}

\subsection{No time delay measurements}\label{sec:ntdm2}
Since the cross-section for multiple imaging is proportional to $v^4$,
the probability for multiple sources to be strongly lensed by a galaxy
size halo is small. However, in sufficiently deep exposures of any
given massive clusters, we expect to find a number of sources being
lensed into multiple images. The vast majority of these sources will
be galaxies in which case we will not be able to measure the time
delays between images. Also, for massive cluster size lenses, time
delays are generally too long to be practically observable even for
variable sources. [The cross-section for two images with brightness
ratio less than $r$ is $\propto\left (\frac{r-1}{r+1}\right
)^2$. Thus, we expect to observe relatively few systems with $r\sim 1$
and small time delays $\Delta t$, see Eq.~(\ref{eq:deltr}).]

\begin{figure}
  	\begin{center} \epsfxsize=\textwidth
  	\epsffile{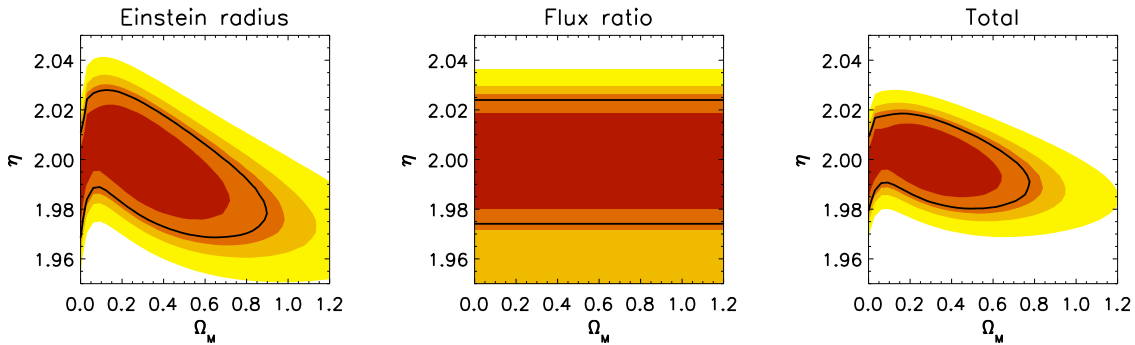} \caption{Results from fitting the
  	image positions for 16 sources at regular intervals between
  	$z=1$ and $z=4.75$ behind a SIS cluster at $z=0.5$ with
  	$v=1000$ km/s. The uncertainty in the oberved image positions
  	is $\sigma_{\theta} \sim 0.2''$.} \label{fig:einstein}
  	\end{center}
\end{figure}

The Einstein radius is a function of both the halo slope and
cosmological distances, see Eqns.~(\ref{eq:theta_E}) and
(\ref{eq:theta_SIS}) and in cases where we have multiple images of
sources at different redshifts, we can use the observed image
positions to constrain $\eta$ and $\om$. A sample of 28 multiply
lensed galaxies with redshift measurements or reliable photometric
redshifts behind Abell 1689 gives a constraint $\om + \ola <1.2$,
i.e.~the data is in accordance with an Einstein-de Sitter universe
with $\om =1$. The constraining power of the sample is limited by the
low redshift ($z\sim 0.18$) of the cluster
\cite{Broadhurst2005b}. However, an analysis of four multiple image systems 
in Abell 2218 (also at $z\sim 0.18$) gives $0<\om <0.3$ assuming a
flat universe, ruling out an Einstein-de Sitter universe at $5\sigma$
\cite{Soucail2004}. The difference in the claimed accuracy of the
results warrants further analysis of the sensitivity of the method on
the model assumptions and uncertainties. This question has been
investigated in great detail in, e.g.~Ref.~\cite{Dalal2005}. Here, we
investigate the sensitivity of the bound on $\om$ on the slope of the
dark matter halo and the size of the observational uncertainties. We
simulate a gravitational lens system consisting of a massive SIS
cluster at $z=0.5$ with $v=1000$ km/s. We then distribute 16 sources
at regular intervals between $z=1$ and $z=4.75$ with random source
positions (inside the cross-section for multiple imaging). Note that
this analysis is very primitive in the sense that we have not included
any of the complexities of real cluster lenses such as substructure
and departures from spherical symmetry. However, this simplicity
allows us to compare the efficiency of the method compared to the ones
described in Sec.~\ref{sec:imis}. The accuracy of our results is
completely determined by the observational error in image position. In
Fig.~\ref{fig:einstein}, results for $\sigma_{\theta} =0.2''$ is
shown, yielding error contours comparable to the ones in
Ref.~\cite{Soucail2004} for $\om$. Setting $\sigma_{\theta} =0.5''$
basically leaves $\om$ unconstrained while $\sigma_{\theta} =0.02''$
yields $\om =0.3\pm 0.04$ and constrains $\eta$ at the per mille
level\footnote{Of course, at this level of precision, the lens model
is too crude.}. Results for $\eta$ are very sensitive to the exact
source positions since large values of $r$ dramatically improve the
limits on $\eta$, see Sec.~\ref{sec:summary}.

Our analysis shows that for a massive cluster with of the order 20
multiply imaged background sources at a large redshift range, we need
the image positions determined to an accuracy of better than
$\sigma_{\theta} \sim 0.2''$ to give useful bounds on $\om$. This
number can be relaxed if the number of sources is larger. The result
is insensitive to any prior information on $\eta$ and thus on the
error in the observed flux ratios (assumed here to be $\sigma_r/r =
0.5$).

\subsection{Time delay measurements}\label{sec:tdm2}
\begin{figure}
  	\begin{center} \epsfxsize=\textwidth \epsffile{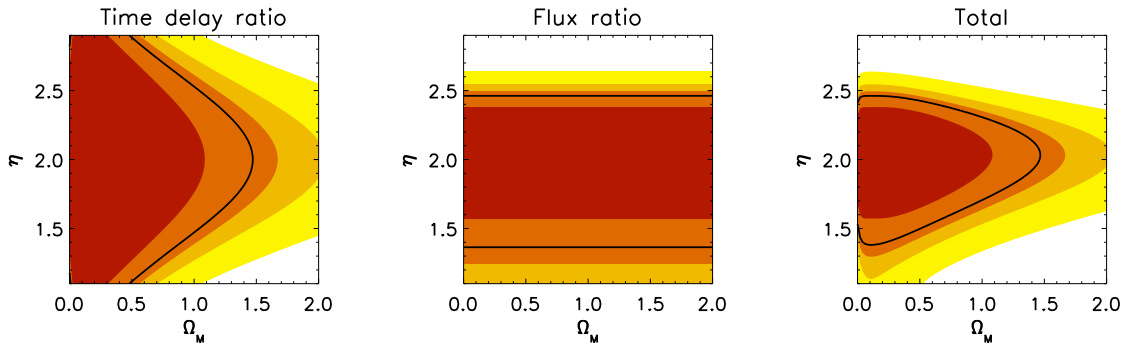}
  	\caption{Results from the time delay ratio (left panel),
  	flux ratios (middle panel) and the combined results (right
  	panel) for a SIS lens at $z_l=0.5$ with a velocity dispersion
  	of $v=300$ km/s and two multiple imaged quasars at $z_{s,a}=1$
  	and $z_{s,b}=3$ at angular source positions $\beta_a=0.5''$
  	and $\beta_b=1.0''$.} \label{fig:bt2} \end{center}
\end{figure}

As explained in Sec.~\ref{sec:ntdm2}, we expect to observe very few
lens systems with multiple images of several variable or transient
sources with time delays short enough to be practically
measureable. Nevertheless, we here explore possible constraints from
such systems. In order to have a non-negligible cross-section for
multiple imaging but reasonable time delays, we imagine lensing from a
very heavy galaxy or a small group of galaxies. Our default system is
a SIS lensing halo at $z_l=0.5$ with a velocity dispersion of $v=300$
km/s. We imagine two multiply imaged quasars at $z_{s,a}=1$ and
$z_{s,b}=3$ at angular source positions $\beta_a=0.5''$ and
$\beta_b=1.0''$. The assumed errors are the same as in
Sec.~\ref{sec:errors} except that we set the error in the time delays
to $\sigma_{\Delta t}=2.0$ days, corresponding to typical errors for
time delays between quasar images \cite{Kochanek2004}. For such a
configuration, we have a time delay for the images of source ``a'' of
$\Delta t_a=148$ days and a flux ratio of $r_a=2.6$. The corresponding
numbers for source ``b'' is $\Delta t_b=296$ days and $r_b=3.4$. We
derive constraints on $\eta$ and $\om$ from the observed flux ratios
and the time delay ratio (c.f.~Sec.~\ref{sec:matter}), see
Fig.~\ref{fig:bt2}. The error budget is equally shared between the
error in the time delay measurements and the observational error in
the image positions of $\sigma_{\theta} =0.01''$. Therefore does our
results not improve significantly for transient sources like SNe or
gamma ray bursts where more accurate time delay measurements are
possible, unless $\sigma_{\theta}$ can be significantly reduced. Note
that the shape and size of the contours in the $[\eta,\om]$-plane is
very dependent on the exact configuration of the lens system. In
general, results benefit from a large redshift separation for the
sources since the time delay ratio is proportional to $(D_{ls}/D_s)_a/
(D_{ls}/D_s)_b$. However, even in the lucky case of finding a lensing
system with two well-observed multiply imaged sources with time delays
determined to good accuracy, we can only hope for very rough
constraints on $\eta$ and $\om$.

\section{Summary and discussion}\label{sec:summary}
We have analyzed and compared how measurements of the image positions,
flux ratios and time delays in different types of strong lensing
systems will allow for the determination of the matter distribution in
galaxies, as well as global cosmological parameters such as the Hubble
constant $h$ and the matter density $\om$. While CMB and LSS data
probe properties of dark matter on large scales, measurements of
galactic halos can probe the small scale properties of dark
matter. Furthermore, gravitational lensing mainly probes halos at
relatively high redshift where few methods for measuring galaxy
density profiles are available.
 
First, we investigated different uses of a data set consisting of a
large number of lens systems, each with a multiply imaged source with
the time delay measured to high precision. Such a data set can
constrain the slope of dark matter halos $\eta$ at the per cent level
and provides an independent test of the values of $h$ and
$\om$. Combining the results from the time delay and flux ratio
measurements, we are able to make a determination of $h$ within 10\,\%
and $\eta$ within 1\,\% ($2\sigma$). 

Factoring out the Hubble constant by studying time delay ratios, we
are able to rule out $\om=1$ at high confidence (assuming a flat
universe). This bound is independent of the slope of the halos and the
flux ratio measurements.

Second, we compared the results obtained for multiple systems with
single sources with the use of lens systems with several multiply
lensed sources in a single lens. For massive cluster size lenses, we
expect to find several multiply imaged galaxy sources. The position of
these images can provide a sanity check on $\om$. The error budget is
dominated by the observational error in image position that need to be
of the order or less than $\sigma_{\theta} \sim 0.2''$ to be able to
rule out an Einstein-de Sitter universe. Systems with multiply imaged
sources with measurable time delays will be rare and not very useful
for the purposes discussed in this paper.

In the case of a large sample of individual multiple image systems,
the error budget in the time delay analysis is dominated by the
observational time delay error. The flux ratio constraints are quite
robust to an increase in the (gaussian) error size.  Increasing the
errors by a factor of 3 (to 150\,\%), we are still able to obtain
$\eta =2\pm0.05$ ($2\sigma$). From Eq.~(\ref{eq:rq}), it is evident
that systems with large flux ratios (i.e.~high $q$) are very important
when constraining galaxy density profiles since the flux ratio in
these systems are more sensitive to changes in $\eta$. If such systems
are absent from the sample, the error on $\eta$ will increase
significantly. Also, increasing the size of the external shear by a
factor of ten causes a systematic bias in the determination of $\eta$
of $\eta=2\rightarrow 1.65$. It should be noted that the use of flux
ratio measurements is potentially very problematic because of large
uncertainties in the microlensing magnification of individual
images. Since the errors are probably not gaussian, additional
information on the probability distribution of the microlensing
magnifications is most likely needed in order to be able to use the
flux ratio information in an unbiased way. On the other hand,
observations of the microlens magnification distribution is
potentially a powerful probe of the small scale structure in galaxy
halos.

A possible complication is that our results for individual multiple
image systems are based on the assumption that we can define average
properties of lensing halos, i.e.~the value of $\eta$, while N-body
simulations indicate large dispersions in the individual halo
properties
\cite{Navarro1997,Ghigna2000,Navarro2004,Fukushige2004,Graham2005}.
Another caveat is that real lensing systems are too complicated to
allow for an analysis as simplistic as in this paper. However, we are
able to constrain $\eta$ and $h$ even with a quality factor $f$ as low
as $0.05$ \footnote{Note however that $\om$ is left more or less
unconstrained for such low values of $f$.}. If it is possible to
retain a large number of lens systems, properties like redshift
evolution of density profiles, variations in density profiles with
galaxy luminosity, colour and many other important parameters can be
measured by dividing the lens systems into different categories (in
which also the scatter in individual halo properties may be smaller).

Undoubtedly, modelling of real future lens systems will use more
complicated lens models and observational information than the one
used in this paper. Nevertheless, we believe that our simple model is
able to show the general parameter dependencies and approximate
confidence contours that can be obtained with future strong lensing
data. Since our results are fairly robust to changes in the quantity
and quality of the observational data, we conclude that strong lensing
should be able to give useful constraints on the matter distribution
in galaxies and galaxy clusters as well as complement other
cosmological probes of the Hubble parameter and the matter density.

\section*{Acknowledgments}
The authors acknowledge helpful comments from an anonymous referee. EM
would also like to thank the Royal Swedish Academy of Sciences for
finacial support.

\section*{References}

\appendix\label{app}
\section{Ellipticity and external shear}
We modify our simple simulated lensing systems by adding the effects
of ellipticities and shear.  In order to treat possible non-sphericity
of the lens systems we add a quadrupole term to the projected
potential, $\Psi$,
\begin{equation}
	\Psi = \Psi_0 \left(1+a \cos 2\phi\right),
\end{equation}
where $\Psi_0$ is the spherical potential given in Eq.~(\ref{eq:psi})
and $\phi$ is the angle of the image relative to the quadrupole axis.
If the system is assumed to have relatively small ellipticity, the
angle of the second image can be written as a function of the first
\begin{equation}
	\phi_2 = \phi_1 + \pi - \delta,
\end{equation}
where $\delta$ is a small parameter (and explicitly zero for spherical
systems).

Most lens systems are embedded within an external potential which
gives rise to an additional shear contribution. We write the combined
potential as
\begin{equation}
\Psi = \Psi_0 \left(1+a \cos 2 \phi\right) + \gamma_{\rm ext} r^2
\cos[2(\phi-\phi_0)],
\end{equation}
where $\phi-\phi_0$ is the angle between the image position and the
quadrupole axis of the external potential. Since $\gamma$ should be
small in order to justify treating the external potential as a
quadrupole, we derive expressions for $\Delta t$ and $r$ which are
valid to second order in $q$, $\delta$ and $\gamma$ for arbitrary
values of $\phi_0$. The expressions are given in
Ref.~\cite{Mortsell2005}. The second order Taylor expanded expressions
are excellent approximations for small $q$ and/or $\eta\sim 2$. In
practice however, we have used the full expressions for the time
delays and the flux ratios. Since $\eta\sim 2$ in our simulations, the
use of Taylor expansions gives close to identical results.

\end{document}